\documentclass[twocolumn,showpreprint,preprintnumbers,amsmath,amssymb]{revtex4}
\usepackage{amsmath} 
\usepackage{amssymb}
\usepackage{epsf}
\usepackage{epsfig}
\usepackage{graphicx}% Include figure files
\usepackage{dcolumn}% Align table columns on decimal point
\usepackage{bm}% bold math
\usepackage{psfrag}
\newcommand{\ud}{\mathrm{d}}

\begin{document} 
\preprint{to be submitted to {\it Appl. Phys. Lett. 2007}}

%\twocolumn[\hsize\textwidth\columnwidth\hsize\csname@twocolumnfalse\endcsname 
\title{Sub-Threshold Field Emission from Thin Silicon Membranes}

\author{Hua Qin$^{1,*}$, Hyun-Seok Kim$^{1}$, 
Michael S. Westphall$^{2}$, Lloyd M. Smith$^{2}$ \& Robert H. Blick$^{1}$}
\address{
$^{1}$Department of Electrical and Computer Engineering, 
University of Wisconsin-Madison, 
1415 Engineering Drive, Madison, WI-53706, USA \\
$^{2}$Department of Chemistry, University of Wisconsin-Madison,
1101 University Avenue, Madison, WI 53706, USA
}

\date{\today}

\begin{abstract} 
We report on strongly enhanced electron multiplication in thin silicon membranes. 
The device is configured as a transmission-type membrane for electron multiplication.
A sub-threshold electric field applied on the emission side of the membrane 
enhances the number of electrons emitted by two orders of magnitude. 
This enhancement stems from field emitted electrons stimulated by the incident particles,
which suggests that stacks of silicon membranes can form ultra-sensitive electron multipliers. 
\\
\textnormal{$^{*}$Electronic mail: QIN1@WISC.EDU}
\end{abstract}

\maketitle

Particle detectors crucially rely on electron multiplication,  
such as the photo and electron amplifier tubes, cathode ray screens and 
micro channel plates~\cite{bruining-54}. 
The multiplication is conventionally achieved by employing the 
so-called secondary electron emission (SEE)~\cite{bruining-54,modinos-84}, 
where a primary electron (PE) strikes the surface of the detector and
induces the emission of lower energy secondary electrons (SEs). 
The physics behind SEE is found in the various interaction pathways between 
the PEs and the solid, e.g., 
electron-nuclear collisions, electron-electron interactions, 
plasmon and phonon excitations~\cite{egerton-86,see-review}. 
To the emission of SEs usually only those contribute, 
which are excited above the vacuum energy level ($E_{vac}$).  
Consequently, the barrier for releasing secondary electrons is determined by the
difference between the vacuum energy level and the Fermi level: $\Phi=E_{vac}-E_F$, 
the work function. 
The actual figure of merit for detectors, e.g., electron multipliers, 
is the secondary electron yield (SEY), which predominantly relies on the detector material.
Naturally, many efforts focus on achieving high SEY by developing new materials 
with a lower work function, such as most recently doped diamond~\cite{auciello}. 

Furthermore, it has been known that the external electric field 
at the surface of a metal or a semiconductor can significantly modify the potential barrier 
and induce field emission (FE) of electrons~\cite{modinos-84}. 
The studies on field-enhanced SEE can apparently be traced back to the discovery of the 
Malter effect in 1936~\cite{malter-36}. 
Follow-up experiments on SEE revealed variants of this field effect~\cite{pensak-49, jacobs-51}. 
These experiments were conducted on bulk metallic substrates coated with a thin layer of oxide, 
where apparently the oxide layer is the key. 
It is this thin surface layer, which either (a) becomes positively charged 
to induce field emission from the underlying emitter and leads to long persistence times of SEE
or (b) induces avalanches of electrons in the oxide layer.

\begin{figure}[!htp] 
%\vspace{1 cm}
\includegraphics[width=0.47\textwidth]{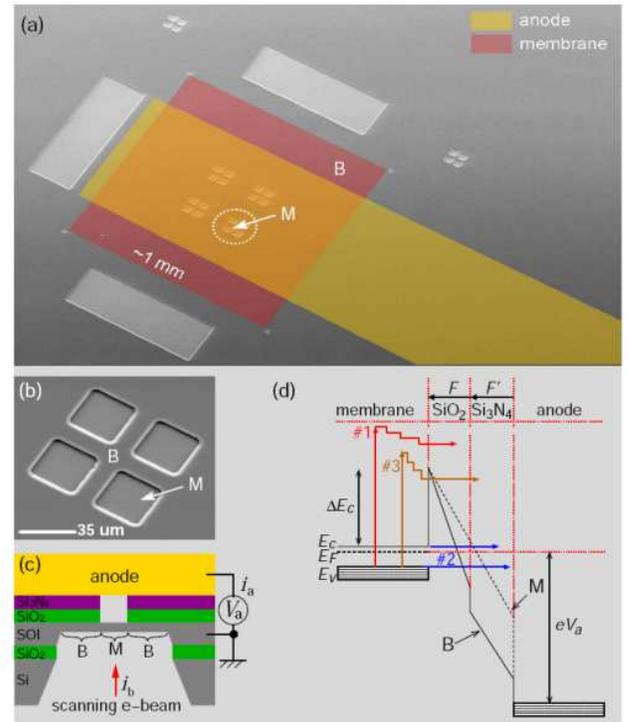}
%\center{Qin {\it et al:~} Figure 1/3}
\caption{
(a) A SEM graph of the device.
(b) A closer view of four silicon membranes (M) 
comparing to triple-layer membrane (B).  
(c) A schematic drawing of the cross section and 
the measurement setup. 
(d) Energy band diagram (not to scale, 
also the band bending of silicon is not shown).
Note the different tunnel barriers for membrane M and B. 
} \label{fig:1}
\end{figure}

Here we present detailed SEE measurements 
on how to dramatically enhance SEY in thin semiconductor membranes by the electric field. 
The field enables sub-threshold electron extraction, increasing the yield by two orders of magnitude. 
We probe this effect in a transmission-type configuration, 
where the back side of the membrane emitter is bombarded by a primary electron beam (e-beam). 
In addition, our technique now allows to probe electronic excitations 
at low energies, which are otherwise not accessible by conventional SEE. 

The device we realized is a thin silicon membrane 
which is capped with two thin layers of silicon oxide and silicon nitride. 
The starting material are silicon-on-insulator (SOI) wafers,  
which consist of a $3.0~\mathrm{\mu m}$ layer of silicon 
on a layer of silicon dioxide ($1.1~\mathrm{\mu m}$). 
Both the SOI and silicon substrate have a crystal orientation of (100) and 
a resistivity of $12~\mathrm{\Omega cm}$, 
corresponding to a {\it n}-type doping level of $3\times10^{14} \mathrm{cm^{-3}}$.  
The Fermi level is about $0.3~\mathrm{eV}$ 
below the bottom of the conduction band at room temperature. 
The SOI wafer was thinned by thermal oxidation to form 
a silicon dioxide layer ($1.0~\mathrm{\mu m}$).
The whole wafer is then capped with a thin layer of silicon nitride ($900~\mathrm{nm}$) 
by using low pressure chemical vapor deposition (LPCVD). 
A square ($1~\mathrm{mm^2}$) membrane of SOI/SiO$_2$/Si$_3$N$_4$ 
triple layer is formed in an anisotropic silicon etching step using 
potassium hydroxide (KOH) solution.   
 
Finally, 16 silicon membranes are formed from this 
triple-layer membrane by carving through the oxide and nitride layers. 
The scanning electron micrograph in Fig.~\ref{fig:1}(a) 
is a top view of the device, 
where the whole triple-layer membrane is seen as a red square 
and in the center are 16 silicon membranes.    
A closer view of four silicon membranes is shown in Fig.~\ref{fig:1}(b). 
Thus two distinct membrane types are fabricated on the same device: 
a pure silicon membrane (M) and a triple-layer membrane (B).  
Each M-type membrane has a dimension of $35 \times 35~\mathrm{\mu m^2}$, 
which is only $2\%$ of the area of the framing B-type membrane. 
The thickness of the M-type membranes is $1.6~\mathrm{\mu m}$. 

We used a modified SEM to conduct the experiments (vacuum of $p \approx 10^{-6}~\mathrm{mbar}$):  
the configuration is sketched in the inset of Fig.~\ref{fig:1}(c), 
where the sample is clamped in a holder 
while the electron-beam is scanned over the backside of the membranes. 
The injected electrons possess energies in the range of $E_p=1-30~\mathrm{keV}$. 
The membranes are connected to ground to avoid charging due to the release of electrons. 
A metallic anode is placed on top of the membrane, covering 
the gold-color area as shown in Fig. 1(a). 
This electrode provides the extraction or retarding voltage ($V_a$) 
for electrons emitted from the membranes. 
By controlling the anode voltage while monitoring the anode current $I_a$, 
SEE ($E \lesssim 50~\mathrm{eV}$) can be differentiated 
from electrons transmitted directly through the membrane ($E \leq E_p$). 
As will be shown below, a further increase in the anode voltage induces FE from the membrane. 

\begin{figure}[!htp] 
%\vspace{1 cm}
\includegraphics[width=0.47\textwidth]{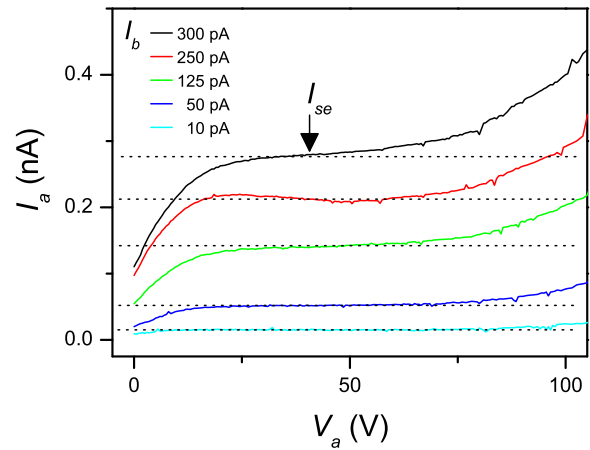}
%\center{Qin {\it et al:~} Figure 2/3}
\caption{
$I_a-V_a$ characteristics reveals conventional SEE 
when $V_a \lesssim 50~\mathrm{V}$. 
The electron-beam is scanned over an area of 
$320 \times 320~\mathrm{\mu m}^2$ 
covering 16 M-type membranes ($E_p = 30~\mathrm{keV}$).
} \label{fig:2}
\end{figure}

%XXXXXXXXXXXXXXXXXXXXXXXXXXXXXXXXXXXXXXXXXXXXXXXXXXXXXXXXXXXXXXXXXXXXXXXXXXX
Energy band diagram of the membranes is schematically shown in Fig.~\ref{fig:1}(d).
The existence of dielectric layers enhances the electric field at the surface of the SOI layer. 
The potential barrier for electron emission in area B is determined 
by the conduction band offset between silicon oxide and silicon ($\Delta E_c$), 
while it is the work function ($\Phi$) of silicon for area M. 
When the backside of the membrane is exposed to an electron-beam, 
SEE can be easily probed by the anode, as shown in Fig.~\ref{fig:2}.
This is indicated by process \#1 shown in the energy diagram in Fig.~\ref{fig:1}(d).
Due to the thinness of the membrane which allows direct electron transmission, 
the anode draws already a finite current even at a negative bias~\cite{bruining-54,qin-07}. 
The increase of $V_a$ up to $+50~\mathrm{V}$ 
allows for a complete collection of both transmitted and true SEs, 
which produces a plateau in the $I_a-V_a$ curve, as shown in Fig.~\ref{fig:2}.   
The height of this plateau normalized by the incident beam current $I_b$
defines the conventional electron yield from SEE: $\gamma_{se}=I_{se}/I_b$, 
which is found to be about $0.95$. 

\begin{figure}[!htp] 
\vspace{1 cm}
\includegraphics[width=0.47\textwidth]{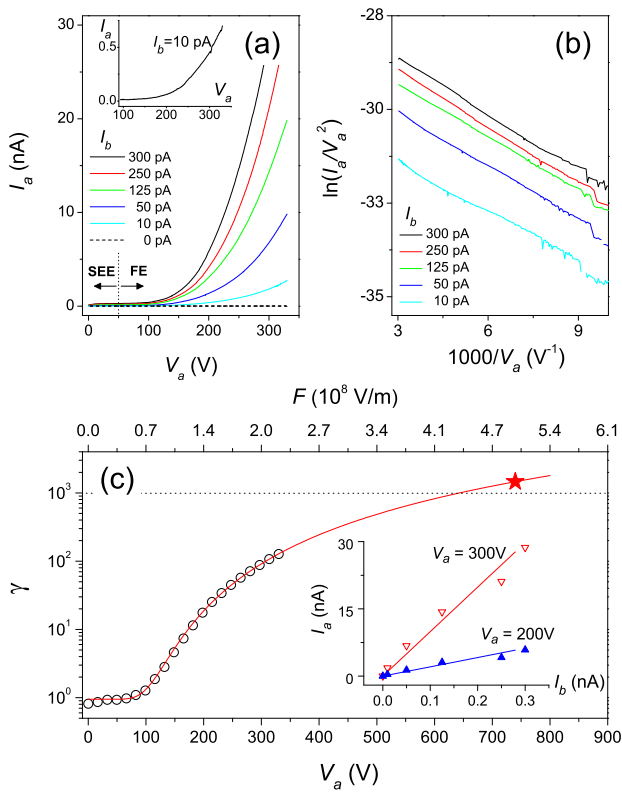}
%\center{Qin {\it et al:~} Figure 3/3}
\caption{ 
(a) Extended $I_a-V_a$ characteristics with the anode voltage up to $350~\mathrm{V}$. 
Other conditions are the same as that shown in Fig.~\ref{fig:2}.
The inset displays electron multiplication from an M-type membrane only. 
(b) The same data shown in (a) presented in a typical FN-plot. 
(c) The total electron yield vs. the anode voltage, revealing a gain of two orders of magnitude
after the transition from SEE to FE. 
The inset shows the linear relationship between $I_a$ and $I_b$ at two different anode voltages. 
} \label{fig:3}
\end{figure}

Without electron-beam excitation on the backside of the membranes, 
electrons in the conduction and valence bands can be emitted through FE, 
shown as process \#2 in Fig.~\ref{fig:1}(d). 
Without injected electrons our device did not show a measurable FE current, 
even when an anode voltage of up to $350~\mathrm{V}$ is applied, 
see the black dashed curve shown in Fig. 3(a). 
This is not too surprising since at $V_a=+350~\mathrm{V}$, 
the electric fields in area B and M  are only of the order of $10^8~\mathrm{V/m}$.  
According to the Fowler-Nordheim (FN) theory, such sub-threshold electric fields 
do not produce a detectable electric current from an emission area of $1~\mathrm{mm}^2$. 
However, when the backside of the membrane 
with an area about $320^2~\mathrm{\mu m^2}$ including 16 M-type membranes 
is exposed to an electron-beam, 
the anode current is significantly increased, as shown in Fig.~\ref{fig:3}(a). 
The electron-beam has an energy of $E_p=30~\mathrm{keV}$ and a beam current as indicated. 
A similar nonlinear curve is also seen in the inset of Fig. 3(a),  
where the electron-beam is fixed on a M-type membrane, i.e. without any dielectric layer.  
This suggests an {\it emission mechanism different from the Malter effect}~\cite{malter-36}
and the other known effects~\cite{pensak-49, jacobs-51} which all rely on 
a thin layer of oxide deposited on the emitter.

The observed strong electron emission with $V_a$ above $100~\mathrm{V}$ 
is further examined along the FN characteristic: $\ln(I_a/V_a^2) - 1/V_a$, 
as shown in Fig.~\ref{fig:3}(b).  
The straight lines in a large range of bias voltage strongly support that 
emission from the membranes is governed by FN tunneling:
$I_a = I_{se}+I_{fe} = \gamma_{se}I_b+A V_a^2 \exp(-B/V_a)$, 
where $I_{se}$ stands for the conventional SEE which saturates around $V_a = 50~\mathrm{V}$ 
(see Fig.~\ref{fig:2}), while the second term is 
the standard FN field emission current ($A$ and $B$ are material dependent constants with 
$B = (512 \pm 31)~\mathrm{V^{-1}}$). 
Most importantly, we found that coefficient $A$ is proportional to the incident electron-beam intensity:
$A = A' I_b =  (5.1 \times 10^{-3}~\mathrm{V^{-2}}) I_b$. 

The total yield can now be expressed as 
$\gamma = I_a/I_b=\gamma_{se}+\gamma_{fe} = \gamma_{se} + A'V_a^2 \exp(-B/V_a)$, 
which can be continuously tuned by the anode voltage. 
As shown in Fig.~\ref{fig:3}(c),  
a yield of 200 is achieved by strong FE at a bias of $350~\mathrm{V}$.  
Below $100~\mathrm{V}$, the yield is reduced to $0.95$ which is determined by conventional SEE. 
The solid curve shown in Fig.~\ref{fig:3}(c) 
is a fit with $A'=(5.2 \times 10^{-3})V^{-2}$ and $B=492~\mathrm{V^{-1}}$, which traces $\gamma$ perfectly. 
In the inset of Fig.~\ref{fig:3}(c) plots are represented for constant yields. 
Here we fixed the anode voltage 
and varied the electron-beam current ($1-300~\mathrm{pA}$).

%Here comes the discussions,
The above results reveal that electrons in a thin silicon membrane can be generated and liberated 
by the combined action of PEs and an external electric field. 
In other words, the external sub-threshold electric field induces field emission of 
those electrons excited by the primaries, which would otherwise remain undetected.
Unlike the Malter effect, a dielectric layer on the surface of the membrane is
not a must in our experiments and no self-sustained emission was observed 
after the scanning electron-beam is turned off~\cite{beamsize}.

%%these are the discussion on possible theory
The main pathway for this sub-threshold field emission which we found here, 
is illustrated as process \#3 shown in the band structures of Fig.~\ref{fig:1}(d). 
the corresponding portion of the emission current should follow 
$I_{fe} \propto \int_{E_c}^{E_c+\Delta E_c} N(E)T(E,V_a) \ud E$, 
where $N(E)$ is the number of electrons excited to the 
energy between $E$ and $E+\ud E$, $T(E,V_a)$ is the tunneling probability. 
The spectrum of possible electron excitations in the thin silicon membrane by the PEs 
should be a continuum, starting from the bottom of the conduction band to the energy level of PEs. 
Electrons at the membrane surface with an energy 
above the tunnel barrier are true SEs since they are free to escape from the membrane.
However, there are many more electrons excited with their energy below the barrier. 
Especially, for electrons with an energy close to the top of the barrier will have a tunneling probability 
as high as $10^{-2}-10^{-1}$. 
%Due to the cascade of scattering, 
%electrons excited well above the barrier but far from the emission surface 
%will loss the energy gradually down to the bottom of the conduction band. 
%As long as they can reach the accumulation region, which is about $10~\mathrm{nm}$ away from the surface, 
%electrons are ready to tunnel from the membrane. 
%%theory stops here

Following this model, 
a yield as high as 1000 could be realized at $V_a \approx 650~\mathrm{V}$ 
or equivalently $F~\approx 4.4 \times 10^8~\mathrm{V/m}$, 
which is still below the dielectric breakdown voltage of silicon dioxide 
($\sim 5 \times 10^8~\mathrm{V/m}$, marked by $\bigstar$  in Fig.~\ref{fig:3}(c)).   
Importantly, such a (flat) membrane as an emitter covered with silicon oxide 
will not sustain a high electric field suitable for  
intrinsic field emission (processes \#2 in Fig.~\ref{fig:1}(d)) 
without triggering a dielectric breakdown in the oxide layer.
By introducing nanopillars as electron emitters onto thin membranes, 
we would be able to achieve an even higher yield and 
observe a transition from stimulated sub-threshold FE to intrinsic FE~\cite{qin-07, kim-07}.

Apart from applications in transmission-type electron multipliers~\cite{wilock-60}, 
this effect also allows for determination of the excitation spectrum, such as, 
phonons and plasmons, both of which have an
energy below the tunnel barrier and are not detectable by conventional SEE. 
Electron emission assisted by these quasiparticles is essential 
for developing particle/radiation detectors with an ultra-high sensitivity and low dark current.  

%conclusions
In conclusion we have demonstrated enhancement of electron multiplication of 
at least two orders of magnitude in a silicon
membrane device. The experimental results suggest that this field-dependent multiplication does not 
require a dielectric layer on the membrane as found in Malter's experiment.  
The multiplication is simply realized by sub-threshold Fowler-Nordheim field emission. 
The device architecture demonstrated allows for 
an easy realization of stacked electron and other particle detectors 
and for probing electronic, photonic, or phononic excitations of thin semiconductor membranes.

%%%%%%%%%%%%%%%%%%%%%%%%%%%%%%%%%%%%%%%%%%%%%%%%%%%%%%%%%%%%%
We like to thank the Wisconsin Alumni Research Foundation for direct support of this work and the National Science
Foundation for partial support (MRSEC-IRG1). 

%%%%%%%%%%%%%%%%%%%%%%%%%%%%%%%%%%%%%%%%%%%%%%%%%%%%%%%%%%%%%

%%%%%%%%%%%%%%%%%%%%%%%%%%%%%%%%%%%%%%%%%%%%%%%%%%%%%%%%%%%%%
%\newpage
%\begin{figure}
%\end{figure}

%%%%%%%%%%%%%%%%%%%%%%%%%%%%%%%%%%%%%%%%%%%%%%%%%%%%%%%%%%%%%
%\newpage


\begin{thebibliography}{10}
\bibitem{beamsize}
It has to be noted that the electron yield was found 
to be independent of the spot size of the electron-beam. 
At $30~\mathrm{keV}$, an electron-beam of $10~\mathrm{pA}$ and 
a spot size of $50~\mathrm{nm}$ inject single electrons one by one into the membrane. 
This suggests that the observed electron multiplication stems from the 
interaction between individual PEs and the thin membrane.
\end{thebibliography}
\end{document}